\documentclass[titlepage]{article}
\usepackage{fullpage}\usepackage{epsfig}\usepackage{cite}\usepackage{ifthen}
\usepackage{times}\usepackage{calc}\usepackage{flafter}

\title{Factors that Affect the Folding Ability of Proteins}
\author{Aaron R. Dinner$^{1,2}$, Victor Abkevich$^1$, 
Eugene Shakhnovich$^{1,2}$ \\ and Martin Karplus$^{1,2,3}$\\ \\
$^1$Department of Chemistry and Chemical Biology\\ 
Harvard University \\ 
12 Oxford St.\\
Cambridge, MA 02138, U.S.A.\\ \\
$^2$Committee on Higher Degrees in Biophysics \\
Harvard University\\ Cambridge, MA 02138, U.S.A.\\ \\
$^3$Laboratoire de Chimie Biophysique\\Institut le Bel\\
Universit\'{e} Louis Pasteur\\ 
4 Rue Blaise Pascal \\
67000 Strasbourg, France \\ \\
Correspondence to: Martin Karplus \\
E-mail: marci@tammy.harvard.edu\\ 
Phone: (617) 495-4108 \\
Fax: (617) 496-3204 \\ \\
Short Title: Factors that Affect Folding Ability
submitted to PROTEINS}

\begin{document}
\maketitle

\pagebreak

\begin{abstract}
The folding ability of a heteropolymer model for proteins subject to 
Monte Carlo dynamics on a simple cubic lattice is shown to be strongly
correlated with the energy gap between the native state and the 
structurally dissimilar part of the spectrum.   We consider a number of 
estimates of the energy gap that can be determined without simulation, 
including the gap in energy between the native and first 
excited fully compact states for sequences with fully compact native states.
These estimates are found to be more robust predictors of folding
ability than a parameter $\sigma$ that requires simulation for its evaluation:
$\sigma = 1 - T_f/T_\theta$, where $T_f$ is the temperature at
which the fluctuation of the order parameter is at its maximum and 
$T_\theta$ is the temperature at which the specific heat is at its maximum.
We show that the interpretation of $T_\theta$ as the collapse transition
temperature is not correct in general and that the correlation between
$\sigma$ and the folding ability arises from the fact that $\sigma$
is essentially a measure of the energy gap.
\end{abstract}

\pagebreak

To function as a protein, a polypeptide must possesses a unique native 
state that is stable at a physiological temperature and be able to find 
that state in a reasonable time (milliseconds to minutes) at that 
temperature in spite of the fact that the number of possible configurations 
prevents it from making an exhaustive search 
(Levinthal paradox) \cite{Levinthal1969}.
For understanding the mechanism of folding and for protein design, 
it is important to determine the properties that distinguish polypeptide
sequences that satisfy these requirements from those that do not.
One approach is to compare sequence attributes with folding ability.
Analytical heteropolymer theory based on a random energy model
suggested that a large gap between the ground (native) state and the states
(folds) that are structurally dissimilar is sufficient to allow a sequence
to adopt a unique, stable structure 
\cite{Bryngelson1987,Bryngelson1989,Shakhnovich1989b}.   
\v{S}ali, Shakhnovich, and Karplus (SSK) demonstrated computationally
that, for a 27-mer random heteropolymer model of a protein subject to 
Monte Carlo (MC) dynamics on a simple cubic lattice,
a large energy gap promotes fast folding as well \cite{Sali1994a,Sali1994b}. 
The energy gap criterion has been confirmed by similar studies by
other authors \cite{Chan1994,Socci1994} and has been used 
in the design of fast folding sequences \cite{Shakhnovich1993a}.  
Moreover, SSK showed 
that, when the ground state is maximally compact, the
gap criterion can be simplified to a consideration of the difference in energy 
between the ground state and the first fully compact ($3\times 3\times 3$) 
excited state \cite{Sali1994a}.  

Based on results obtained with a set of 15-mer and 27-mer sequences, 
Klimov and Thirumalai (KT) argued that folding ability is not determined 
by the energy gap but instead by the parameter $\sigma = 1 - T_f/T_\theta$,
where $T_f$ is the ``folding'' transition temperature (defined as the
maximum in the fluctuation of the order parameter) and $T_\theta$ is the 
``collapse'' transition temperature (defined as the maximum in the specific 
heat) \cite{Klimov1996a,Klimov1996b}.   KT found that $\sigma$ correlates 
positively with the folding time (small $\sigma$ yields fast folding).
In the present paper, we show that the maximum in the specific heat
($T_\theta$) is not the temperature of the collapse transition in general
and that $\sigma$ is essentially a measure of the energy gap.  In light of
these considerations,
we re-examine the results for the sequences studied by KT.
We find that other measures of the energy gap and native state stability, 
such as the Z-score, correlate more strongly with folding ability than does
the parameter $\sigma$.

The specific model is a self-avoiding heteropolymer chain on a
three-dimensional simple cubic lattice.  The energy function is the sum 
over all contacts (non-bonded spatial nearest-neighbors):
\begin{equation}
E = {\sum}_{i<j-1} B_{ij}\Delta (r_{ij} - a)
\end{equation}
where $r_{ij} = |{\bf r}_i - {\bf r}_j|$ is the distance between monomers 
$i$ and $j$ located at ${\bf r}_i$ and ${\bf r}_j$, respectively, and  $a = 1$
is the lattice spacing.
The function $\Delta (r_{ij} - a)$ selects the interacting monomer
pairs; $\Delta(0) = 1$ and 0 otherwise.
The $B_{ij}$ give the specific interaction energies between monomers 
$i$ and $j$; a complete set of $B_{ij}$ defines a sequence. 
The $B_{ij}$ are chosen randomly with Gaussian probability distribution:
\begin{equation}
P(B_{ij}) = \frac{1}{\sqrt{2\pi}B}
\exp\left[-\frac{1}{2}\left(\frac{B_{ij} - B_0}{B}\right)^2\right].
\label{PBij}
\end{equation}
The quantity $B_0$ ($B_0 < 0$) is a mean attraction between monomers
that corresponds to an overall hydrophobic term.
The 200 27-mer sequences studied in \cite{Sali1994a} were generated with
$B_0 = -2.0$ because that value guaranteed that most sequences had maximally
compact ground states which could be found by examining the enumerated ensemble
of 103,346 maximally compact structures \cite{Shakhnovich1990b}.
In the present study, we use the same sequences 
as in \cite{Klimov1996a,Klimov1996b}; there are nine 15-mers with $B_0 = -2.0$, 
thirty-two 15-mers with $B_0 = -0.1$,  two 27-mers with $B_0 = -2.0$, and 
fifteen 27-mers with $B_0 = -0.1$.  KT included
the value $B_0 = -0.1$ because they felt that it produced a more realistic
fraction of hydrophobic interactions.

It should be noted that some of the sequences (KT have not stated which 
ones) were optimized by a Monte Carlo procedure in sequence space 
\cite{Klimov1996b}.  This procedure differs from that used by others 
\cite{Shakhnovich1993a} in that individual elements of the $B_{ij}$ matrix 
were interchanged, rather than entire rows and columns 
(the latter is equivalent to switching residue types in a linear sequence).
This optimization procedure is free to make the native contacts the 
lowest energy ones, so that there is no frustration, and, 
because only odd $|i-j| \ge 3$ can form contacts on a simple cubic lattice,  
to change the distribution of the $B_{ij}$ that actually contribute
to the energy function.   The averages of the $B_{ij}$ with odd $|i-j|$
($\bar{B}_{ij}$) differed substantially from $B_0$ in the 58 sequences
studied; $-0.805 \leq \bar{B}_{ij} \leq 0.690$ 
for sequences with $B_0 = -0.1$ and $-2.413 \leq \bar{B}_{ij}
\leq -1.708$ for sequences with $B_0 = -2.0$.

We follow KT and use as an order parameter the complement of native 
pairwise distances \cite{Klimov1996a,Klimov1996b}:
\begin{equation}
\chi = 1 - \frac{1}{(N-2)(N-1)}\sum_{i\ne j, j\pm 1}\Delta(r_{ij}-r_{ij}^0),
\end{equation}
where $r_{ij}^0$ is the distance between $i$ and $j$ in the ground
state, and we use as a measure of compactness the total number of 
contacts ($C$).  Boltzmann weighted averages of these quantities were 
calculated exactly for the 15-mer (all 93,250,730 self-avoiding conformations 
can be enumerated) and by Monte Carlo simulation for the 27-mer 
(see \cite{Klimov1996b} for the protocol).  Following KT,
the ``folding'' transition temperature ($T_f$)  is taken to be the 
temperature at which the fluctuation in the order parameter $\Delta\chi(T) =
\langle \chi^2\rangle -  \langle \chi\rangle^2$ has a maximum, and the 
``collapse'' transition temperature ($T_\theta$) is taken to be the 
temperature at which the heat capacity $C_v(T) = \Delta E/T^2 = 
(\langle E^2\rangle - \langle E\rangle^2)/T^2$ has a maximum (units are 
chosen to make the Boltzmann constant equal to 1).  
There is one 15-mer sequence with $B_0 = -0.1$
that exhibited two peaks in $C_v(T)$, three that exhibited shoulders
with derivatives close to zero in $C_v(T)$, and one that
exhibited a shoulder with a derivative close to zero in $\Delta\chi (T)$;
in these cases, for consistency with KT, we calculated the corresponding 
transition temperature ($T_\theta$ or $T_f$) by a weighted average of the two 
temperatures corresponding to the zero derivative points (see Eqn.~10 of 
\cite{Klimov1996b}).

\begin{figure}[hbt]
\begin{center}
\epsfig{file=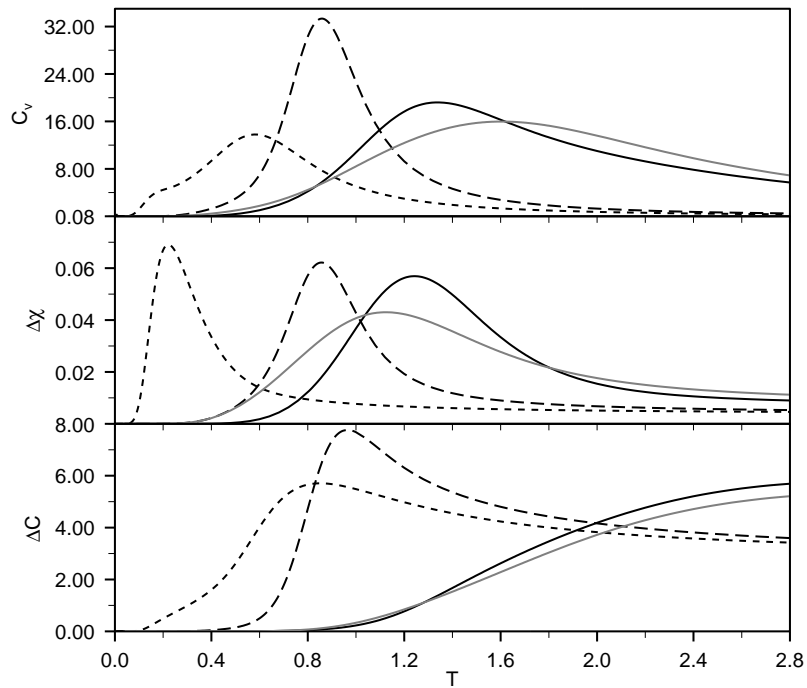, height = \textheight/2 - 2cm} 
\end{center}
\caption{\label{flucs} Exact fluctuations for representative 15-mers: 
$B_0 = -2.0$ sequences (black solid) 91 ($\sigma = 0.071$) and
(gray solid) 95 ($\sigma = 0.299$) and $B_0 = -0.1$ sequences 
(long dashed) 62 ($\sigma = 0.006$) and (short dashed)  5 ($\sigma = 0.621$).
}
\end{figure}

From $T_f$ and $T_\theta$, we can calculate the parameter 
$\sigma = 1 - T_f/T_\theta$, which was argued by KT to determine 
folding ability \cite{Klimov1996a,Klimov1996b}.   To better understand
$T_f$, $T_\theta$, and $\sigma$, we examine the heat capacity ($C_v$), the fluctuation in 
the order parameter ($\Delta\chi$), and the fluctuation in the total 
number of contacts ($\Delta C$) as functions of temperature for several 
representative 15-mer sequences (Fig.~\ref{flucs}).  
When $B_0 = -2.0$, the collapse is determined by the overall attraction
between monomers; $C_v(T)$ reaches its maximum ($T=1.340$ for sequence 91
and $T=1.605$ for 95) at a much lower temperature than does 
$\Delta C(T)$ ($T= 3.160$ and $T\approx 3.320$, respectively).  
When $B_0 = -0.1$, the collapse transition is driven 
by specific contacts so that the folding and collapse transitions are closer 
in $T$.   In the case of sequence 62 ($\sigma = 0.006$), the
peak in $C_v$ ($T=0.860$) is closer to the peak in $\Delta\chi$
($T= 0.855$) than to the peak in $\Delta C$ ($T= 0.965$). 
In the case of sequence 5 ($\sigma = 0.621$), the peak in $C_v$
($T= 0.580$) is closer to the peak in $\Delta C$ ($T= 0.850$)
than to the peak in $\Delta\chi$ ($T= 0.220$), but there is a 
substantial shoulder in $C_v$ at the temperature corresponding to the peak 
in $\Delta\chi$, which is not uncommon for sequences with larger $\sigma$.
Although $T_f$ is clearly associated with the folding transition, 
$T_\theta$ cannot be interpreted as the temperature of the
collapse transition in general.

It is apparent from the above that $\sigma$ must have a different physical 
meaning than that given in \cite{Klimov1996a,Klimov1996b}.  
To better understand $\sigma$, we consider a simple two-state model with
free energy $F_i = E_i - TS_i$ ($i = 1, 2$).
By substitution into $\Delta\chi = \langle\chi^2\rangle - \langle\chi\rangle^2$,
we find
\begin{equation}
\Delta\chi = (\chi_1 - \chi_2)^2\frac{\exp\left(\frac{F_1-F_2}{T}\right)}
{\left[1 + \exp\left(\frac{F_1-F_2}{T}\right)\right]^2}.
\end{equation}
Similarly, 
\begin{equation}
C_v = \left(\frac{E_1 - E_2}{T}\right)^2\frac{\exp\left(\frac{F_1-F_2}{T}\right)}
{\left[1 + \exp\left(\frac{F_1-F_2}{T}\right)\right]^2}.
\end{equation}
To determine the maxima ($T_f$ and $T_\theta$, respectively), 
we take the derivates 
with respect to temperature and set them equal to zero; it is important 
to note that $\chi_1-\chi_2$ and $E_1-E_2$ are temperature dependent.
We solve the resulting equations by expanding
around the true transition point [$T_t = T(F_1  = F_2)$]:
\begin{equation}
\label{expand}
\exp\left(\frac{F_1-F_2}{T}\right)\cong 1 - (E_1 - E_2)\frac{T-T_t}{T_t^2}.
\end{equation}
Substitution of Eqn.~\ref{expand} into $d\Delta\chi/dT = 0$ and $dC_v/dT = 0$
and straightforward rearrangement yield
\begin{equation}
\label{egap}
T_\theta - T_f \propto 1/(E_1 - E_2)^2.
\end{equation}
In other words, $T_\theta$ and $T_f$ deviate from the true transition
($T_t$) in such a way that a measure of their difference ($\sigma$) is 
inversely proportional to the square of the energy gap. 
Although lattice models (particularly the short chains considered
here) deviate from two-state behavior, the simple model suggests that $\sigma$ 
is related to the energy gap.

We now compare the correlation between $\sigma$ and folding ability
to that obtained with other measures of the native state stability.  
These include the transition temperatures themselves ($T_\theta$ and $T_f$) 
and the Z-score of the native state \cite{Shakhnovich1993a}, which more 
directly measures the degree to which a native state is separated from the 
majority of states in the spectrum.  For all sequences, we estimated the 
Z-score by $Z_B = (E_0 - C_0\bar{B}_{ij})/\sigma_B$,
where $E_0$ is the energy of the native state, $C_0$ is the number of 
contacts in that state, $\bar{B}_{ij}$ is the average $B_{ij}$ and 
$\sigma_B$ is the standard deviation of $B_{ij}$.  
For the 15-mers, it was possible to compare this
estimate to the exact Z-score: $Z_E = (E_0 - \bar{E})/\sigma_E$,
where $\bar{E}$ is the unweighted average energy and $\sigma_E$ is 
the standard deviation of energies.  As expected, the two quantities are 
correlated (Tables~\ref{corr} and~\ref{morecorr}). 
For the sequences that have fully compact ground states, the Z-score
is closely related to the energy gap between the ground and 
first excited compact states ($\Delta_{cs}$) \cite{Sali1994a}. 
To compare with \cite{Sali1994a}, we include statistics for $\Delta_{cs}$
for the nine 27-mers with $B_0 = -0.1$ that have $3\times 3\times 3$ ground states.
Inclusion of the six sequences that have non-compact ground states would
be inappropriate, because $\Delta_{cs}$ is unrelated to the Z-score in
these cases ($r_{\Delta_{cs}Z_B} = -0.306$).

The folding ability of a sequence was measured by the mean-first
passage time for finding the ground state (MFPT).  The MFPT was 
calculated from 100 Metropolis Monte Carlo trials with a move set
identical to that used previously
\cite{Sali1994a,Sali1994b,Klimov1996a,Klimov1996b}.
Trials began in a random configuration and were allowed to proceed for 
up to $10^9$ steps.  Following KT, the simulation temperature ($T_s$) was
chosen to yield a constant value of the order parameter:
$T_s = T(\langle \chi\rangle = 0.21)$.  A similar temperature was
used by SSK: $T_x = T(X = 0.8)$, where $X = 1-\sum_i p_i^2$ 
and $p_i$ is the Boltzmann probability of conformation $i$ \cite{Sali1994a}.
Since $T_s$ is determined from an order parameter, it is closely
related to $T_f$, and is thus another measure of the native stability.

\begin{figure}[hbt]
\begin{center}
\epsfig{file=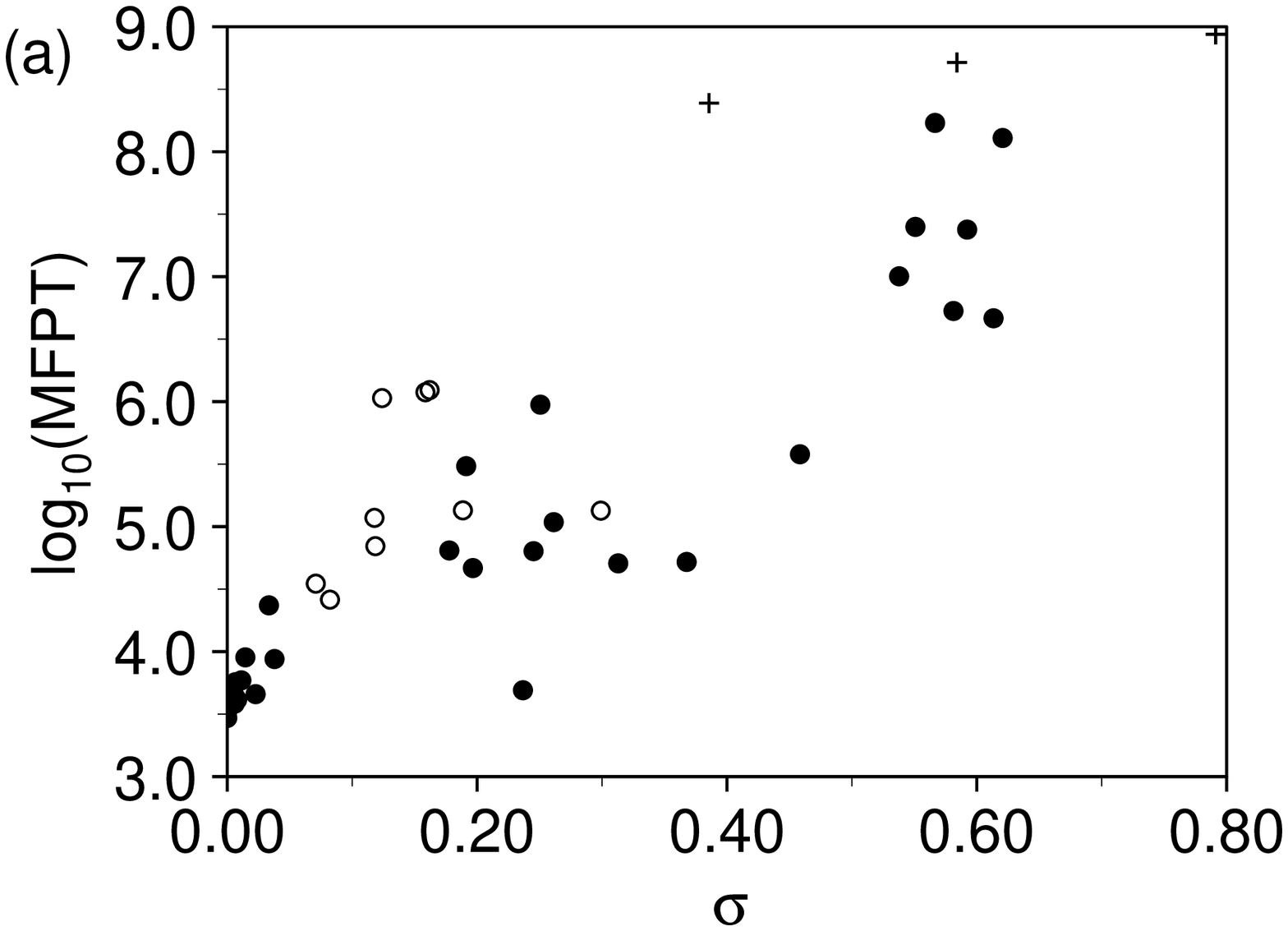, height = \textheight/3 - 2cm}  \\
\epsfig{file=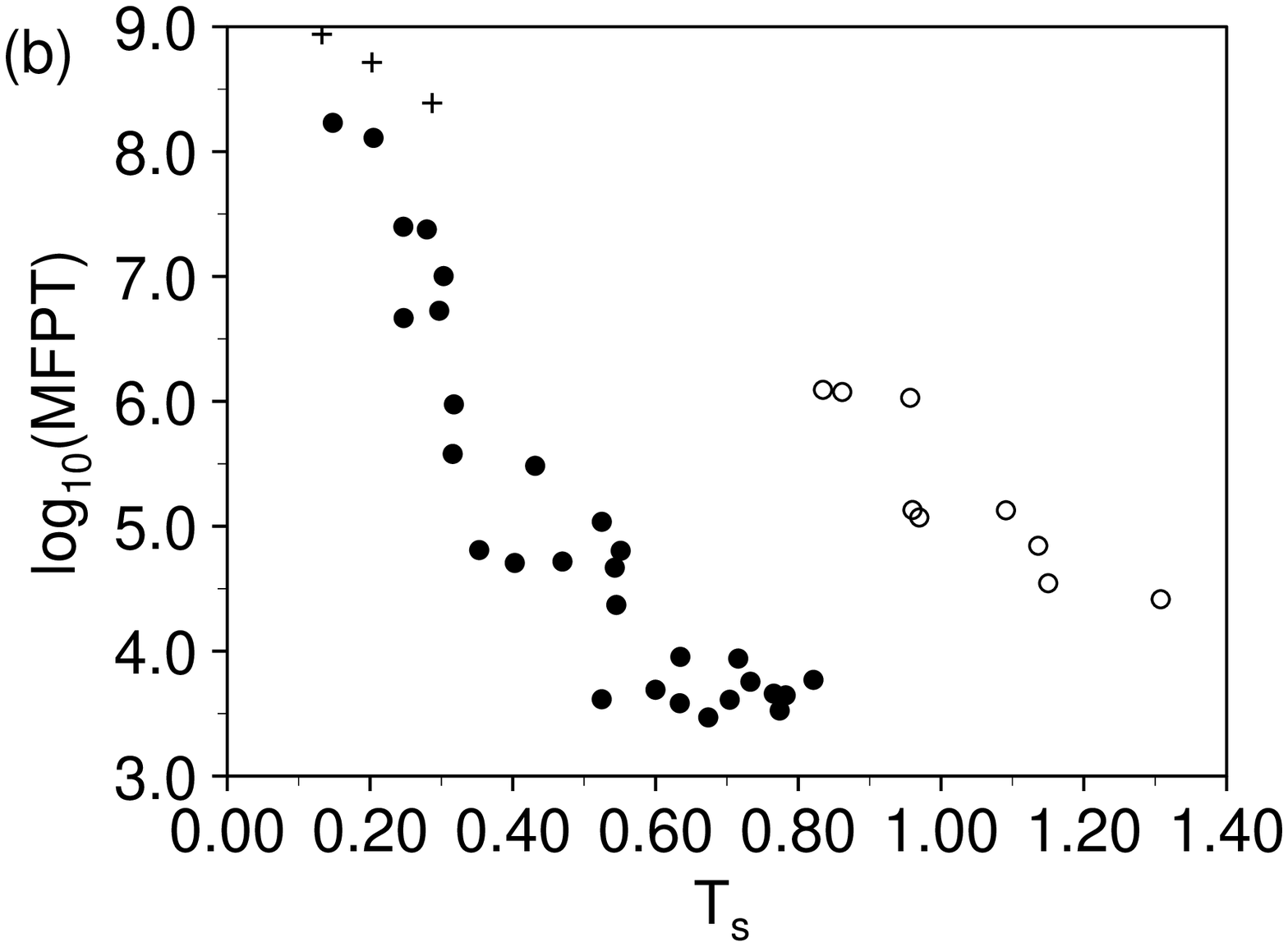, height = \textheight/3 - 2cm}  \\
\epsfig{file=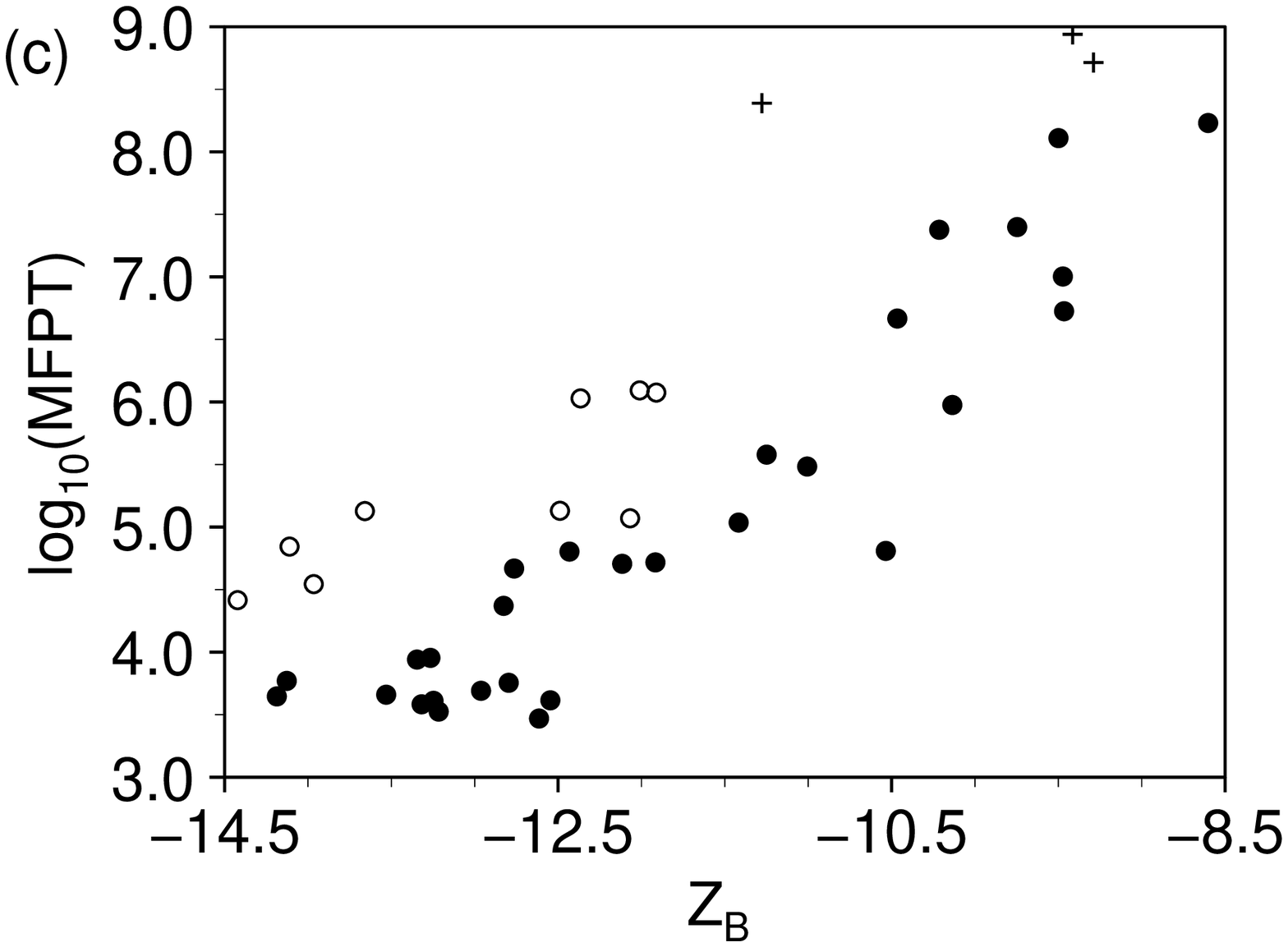, height = \textheight/3 - 2cm} 
\end{center}
\caption{\label{stat} 15-mer dependence of the mean-first passage time on
$\sigma$, $T_s$, and $Z_E$ for sequences with $B_0 = -0.1$ ($\bullet$), those with
$B_0 = -2.0$ ($\circ$), and those that were excluded from the statistics (+) (see
Table~\ref{corr} caption).
For the sequences excluded from the statistics, first passage times of $10^9$ were
substituted for the trials which failed to find the native state.}
\end{figure}

\begin{figure}[hbt]
\begin{center}
\epsfig{file=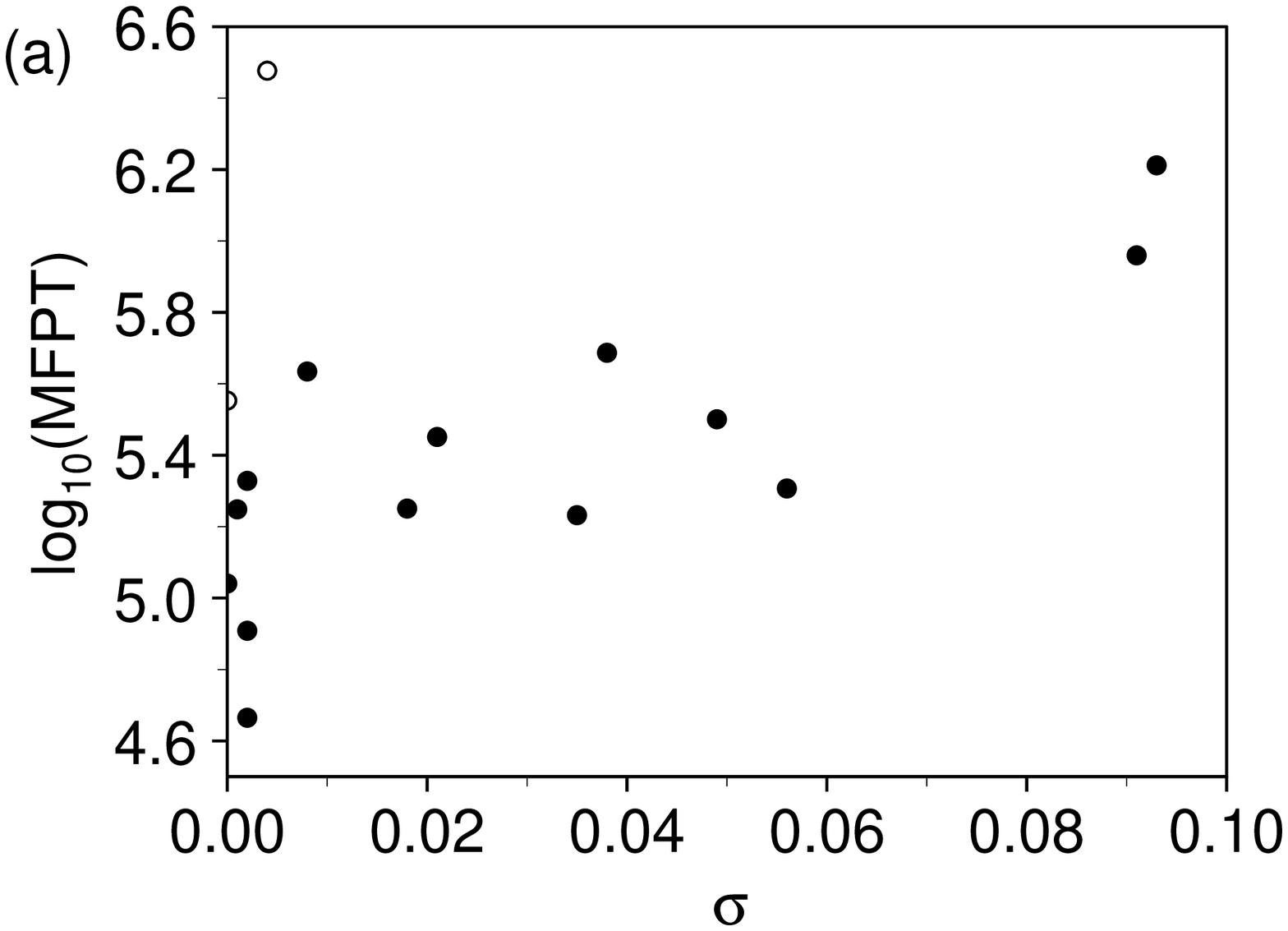, height = \textheight/3 - 2cm}  \\
\epsfig{file=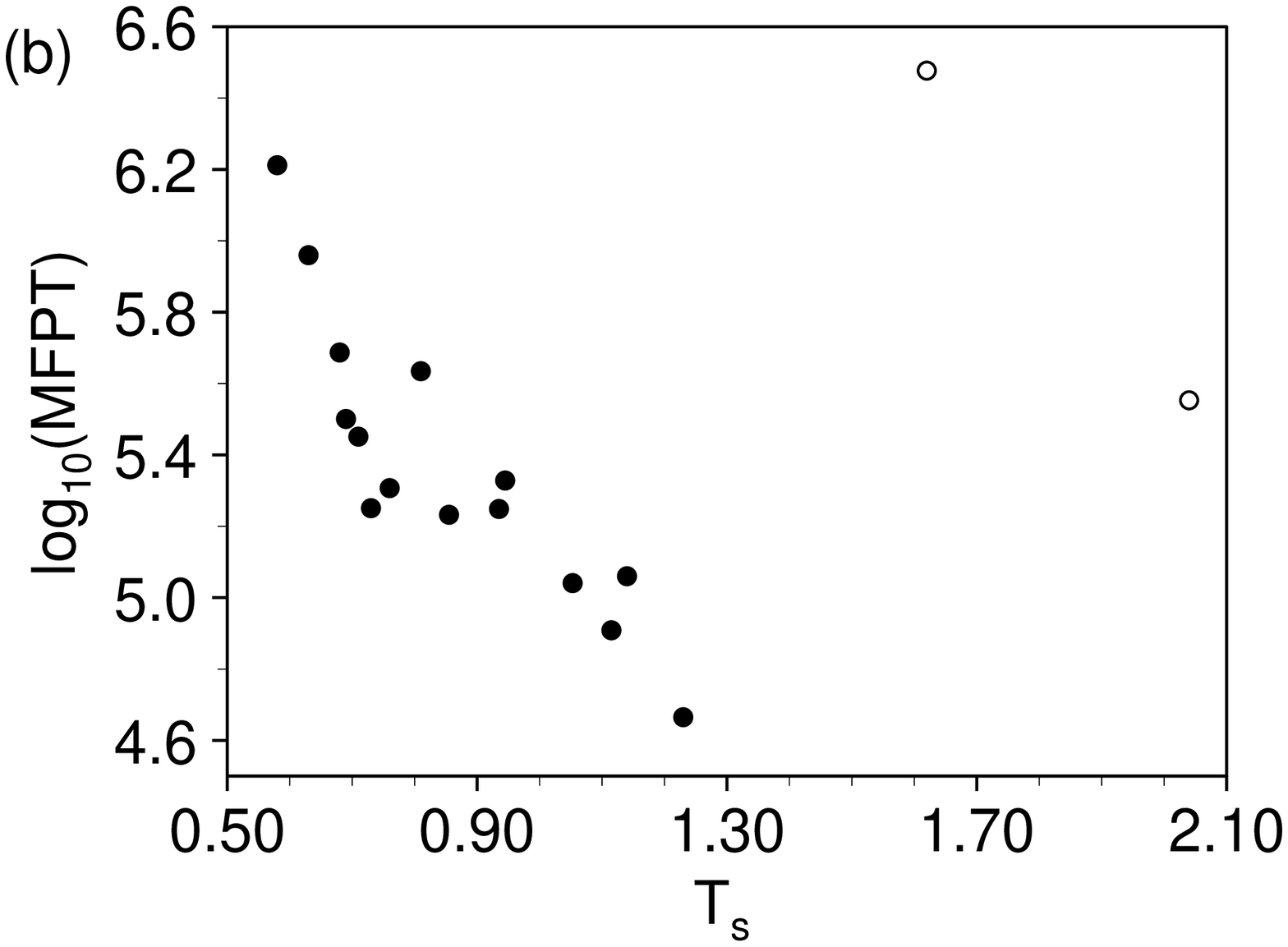, height = \textheight/3 - 2cm}  \\
\epsfig{file=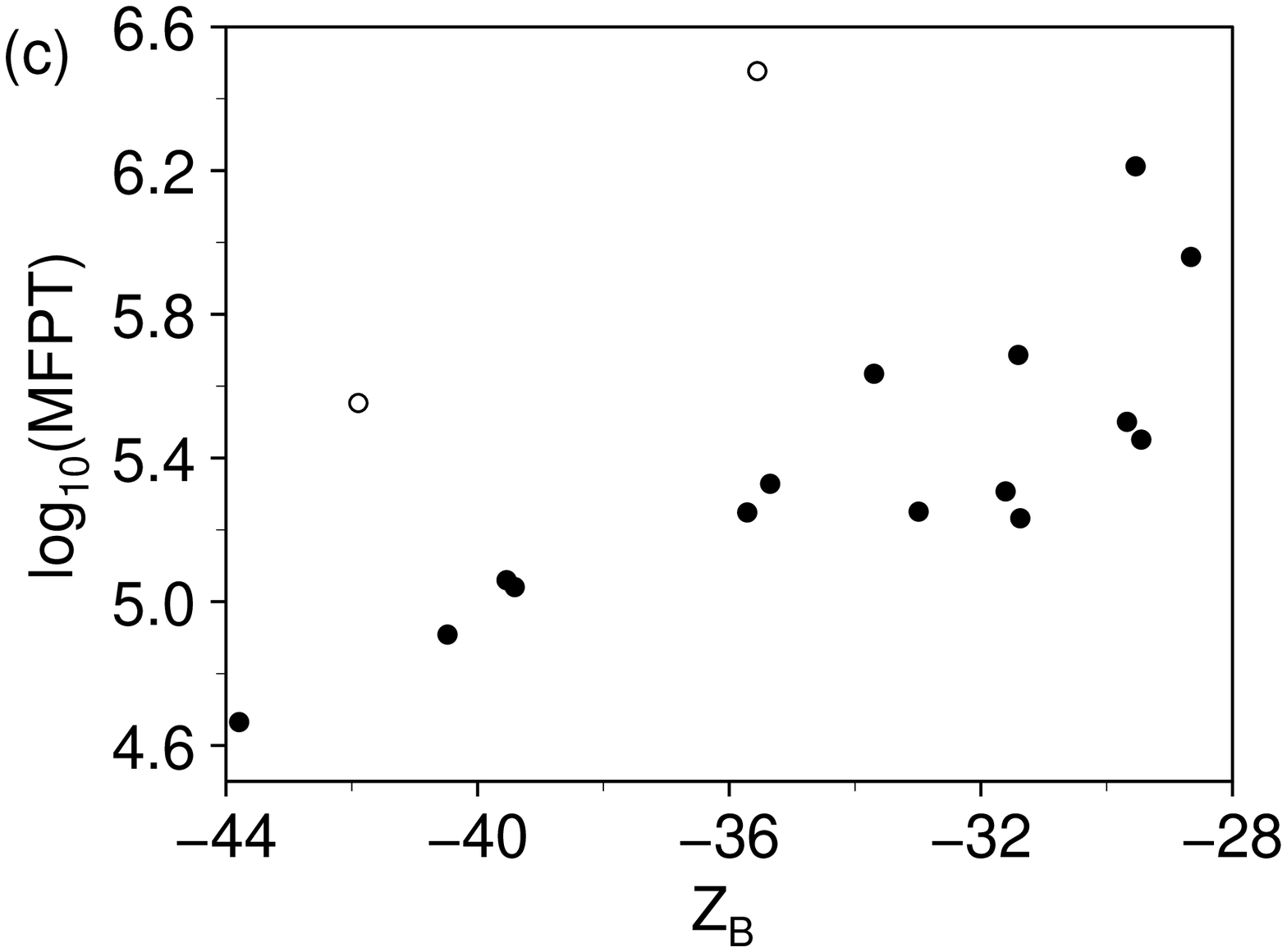, height = \textheight/3 - 2cm} 
\end{center}
\caption{\label{morestat} Same as Fig.~\ref{stat} for 27-mers.}
\end{figure}

The logarithm of the MFPT is plotted against $\sigma$, $Z_B$, and $T_s$ 
for the 15-mers (Fig.~\ref{stat}) and the 27-mers (Fig.~\ref{morestat}). 
Overall, Figs.~\ref{stat}a and \ref{morestat}a are in good agreement with
Figs.~2 of \cite{Klimov1996b} and 11 of \cite{Klimov1996a}.
Pearson linear correlation coefficients (Tables~\ref{corr} and~\ref{morecorr})
are calculated separately for $B_0 = -0.1$ and $B_0 = -2.0$ since these 
represent different solvent conditions.  Although there is a high 
correlation between $\sigma$ and MFPT for sequences with $B_0 = -0.1$, 
the correlation is low for those with $B_0 = -2.0$.   In contrast, 
$Z_B$ and $T_s$ both do well and are comparable for both values of $B_0$.   
For the 27-mers, the most sensitive measure of folding ability is clearly 
$T_s$; these results are in agreement with \cite{Sali1994a}, 
where it was found that a temperature at which the order parameter had 
a particular value ($T_x$) yielded the best predictivity of folding ability.
For the sequences with compact ground states, $\Delta_{cs}$ correlates
more strongly with folding ability than does any other measure, particularly
$\sigma$ (Table~\ref{morecorr}).

The correlation between $\Delta_{cs}$ and folding ability
is obscured in the studies by KT (Fig.~22 of \cite{Klimov1996b})
because they include sequences with native states that are not fully compact,
even though it is clearly stated in \cite{Sali1994a} that doing so is
inappropriate (see also above).   Instead, they compare folding 
ability to the gap between the the native state and the first excited 
state from the complete conformational ensemble ($\Delta$) (Fig.~20 of 
\cite{Klimov1996b}).  Typically, the first {\it non-compact} excited state 
is a tail flip, and thus that gap need not be correlated with either 
$\Delta_{cs}$ or the Z-score. 
Moreover, it should be noted that, even if the 
correct gap ($\Delta_{cs}$) is used with the appropriate sequences
(those with fully compact native states), it is incorrect to ``normalize'' 
by the simulation temperature ($T_s$) as KT have done in Fig.~3 of 
\cite{Klimov1996a} and Fig. 21 of \cite{Klimov1996b}.
Because $\Delta_{cs}$ and $T_s$ are closely related, they both
increase as the magnitude of the Z-score increases. 
Consequently, their ratio, $\Delta_{cs}/T_s$, is not expected to
correlate with stability or folding ability.  The calculations 
presented here are completely consistent with those of
KT \cite{Klimov1996a,Klimov1996b}; it is only the interpretation that
differs.

In the study by KT and the present one, the simulation temperature 
differs for each sequence.  This provides a physically meaningful 
approach since it is appropriate to compare the folding of sequences under 
conditions of corresponding stability;  a protein sequence must not only 
be able to find its native state but it must be dominantly populated
at equilibrium \cite{Sali1994a}.  It has been 
suggested that the correlation between folding ability and the energy
gap found in \cite{Sali1994a} and, by inference, here derives from the
variation in simulation temperature \cite{Unger1996}.  
To address this concern, we performed 
simulations for the 15-mers at three sets of constant temperatures
$T = 0.8, 1.0$, and 1.2 (Table~\ref{constt}).  Although the correlations 
are somewhat reduced, they remain highly significant.  In particular, the 
correlation between the logarithm of the MFPT and the exact Z-score ($Z_E$), 
the most direct measure of the energy gap, is essentially unchanged.

Thus, for the same 58 sequences as were studied in 
\cite{Klimov1996a,Klimov1996b}, it is evident that, although $\sigma$ 
is well correlated with the folding ability for sequences with small $B_0$ 
($B_0 = -0.1$), it is not for sequences with large $B_0$ 
($B_0 = -2.0$).  The difference between the two sets of sequences stems
from a difference in the dependence of $\sigma$ on the Z-score (and hence
$T_s$), which correlates well with folding ability for both values of 
$B_0$ and different simulation temperatures; the relationship between 
$\sigma$ and the energy gap (Eqn.~\ref{egap}) breaks down for sequences with 
$B_0 = -2.0$ because they deviate strongly from two-state behavior
due to the large separation in temperature of the collapse and folding 
transitions.  Moreover, $Z_B$, unlike $T_s$ and $\sigma$, can be calculated
without either explicit enumeration of all the conformations or 
Monte Carlo simulation.  Consequently, $Z_B$ is not only more practical for
protein design, but it is a true predictor of folding ability.

The correlation of the folding rate with the energy gap can be understood in 
terms of its effect on the energy surface.  For random 27-mer sequences,
SSK found that folding proceeds by a fast collapse to a semi-compact random 
globule, followed by a slow, non-directed search through the ($\sim 10^7$)
semi-compact structures for one of the ($\sim 10^3$) transition states that 
lead rapidly to the native conformation \cite{Sali1994b}.  A large energy gap
results in a native-like transition state that is stable at a 
temperature high enough for the folding polypeptide chain to overcome barriers 
between random semi-compact states.   As the energy gap increases 
to the levels obtainable in designed sequences,
the model exhibits Hammond behavior \cite{Hammond1955}
in that the fraction of native contacts required in the transition
state from which the chain folds rapidly to the native state decreases;
random sequences with relatively small gaps must form 80\% of the native
contacts \cite{Sali1994b}, while designed sequences with large 
gaps need form only 20\% \cite{Abkevich1994a,Dinner1996}.  
This decrease in the number of native contacts in the transition state
as the energy gap increases is consistent with the the behavior of
homogeneous systems that coalesce into droplets.  Strengthening the 
interparticle interactions accelerates the process by lowering the 
energy of the transition state and decreasing its size \cite{Oxtoby1988}. \\

\noindent {\Large\bf Acknowledgements} \\

We thank Yaoqi Zhou for helpful discussions and 
KT for making their sequence data available to us.
ARD is a Howard Hughes Medical Institute Pre-Doctoral Fellow.
ES was supported in part by a grant from the National Institute of
Health (GM~52126), and MK was supported in part by a grant from
the National Science Foundation.  

\pagebreak

\bibliography{kt-latt}
\bibliographystyle{unsrt}

\pagebreak

\begin{table}[hbt]
\begin{center}
\begin{displaymath}
\begin{array}{lrrrrrrr}
\hline\hline
& \log_{10}({\rm MFPT}) & T_s & T_\theta & T_f & \sigma & Z_B & Z_E \\ \hline
\log_{10}(MFPT) &      & -0.886 & -0.699 & -0.838 &  0.295 &  0.825 &  0.684 \\
T_s           & -0.910 &        &  0.671 &  0.813 & -0.302 & -0.949 & -0.576 \\
T_\theta      & -0.645 &  0.802 &        &  0.765 &  0.249 & -0.661 & -0.470 \\
T_f           & -0.896 &  0.978 &  0.838 &        & -0.432 & -0.728 & -0.619 \\
\sigma        &  0.928 & -0.899 & -0.526 & -0.897 &        &  0.187 &  0.269 \\
Z_B           &  0.933 & -0.944 & -0.731 & -0.918 &  0.874 &        &  0.682 \\
Z_E           &  0.801 & -0.879 & -0.810 & -0.880 &  0.737 &  0.830 &        \\
\hline\hline
\end{array}
\end{displaymath}
\end{center}
\caption{\label{corr} 
Pearson linear correlation coefficients ($r_{xy}$)
between measures of stability and the logarithm of the mean first
passage time for simulations at $T = T_s$ taken pairwise for  
15-mers with $B_0 = -2.0$ (above) and with $B_0 = -0.1$ (below).
$r_{xy} = \sigma^2_{xy}/\sigma_x \sigma_y =
\sum_i^n (x_i - \langle x\rangle)(y_i - \langle y\rangle)/ \surd
[\sum_i^n (x_i - \langle x\rangle)^2 \sum_i^n (y_i - \langle y\rangle)^2]$.
A perfect correlation has $r_{xy} = 1$, a perfect anti-correlation
has $r_{xy} = -1$.  
Three of the 15-mer sequences with $B_0 = -0.1$ repeatedly 
did not succeed in folding in the allotted time, and so were excluded from the 
statistical analysis.
These sequences  fail to find the native state even though the number of MC
steps is much larger than the number of possible conformations because the
simulation temperature is low, so that most steps are rejected.}
\end{table}

\begin{table}[hbt]
\begin{center}
\begin{displaymath}
\begin{array}{lrrrrrrr}
\hline\hline
& \log_{10}({\rm MFPT}) & T_s & T_\theta & T_f & \sigma & Z_B \\ \hline
\log_{10}(MFPT) &        &-0.881 &-0.826 &-0.859 & 0.799 & 0.829 \\
T_s             & -0.854 &       & 0.934 & 0.961 &-0.785 &-0.965 \\
T_\theta        & -0.864 & 0.998 &       & 0.994 &-0.621 &-0.889 \\
T_f             & -0.857 & 0.998 & 1.000 &       &-0.702 &-0.917 \\
\sigma          &  0.509 &-0.837 &-0.821 &-0.831 &       & 0.752 \\
Z_B             &  0.890 &-0.971 &-0.970 &-0.970 & 0.789 &       \\
\Delta_{cs}   & -0.884 & 0.686 & 0.691 & 0.685 &-0.407 &-0.769 \\
\hline\hline
\end{array}
\end{displaymath}
\end{center}
\caption{\label{morecorr} 
The same as Table~\ref{corr} for 27-mers with $B_0 = -0.1$.
All (above) and those that have fully compact ground states (below).}
\end{table}

\begin{table}[hbt]
\begin{center}
\begin{displaymath}
\begin{array}{lrrrrrr}
\hline\hline
T & \multicolumn{2}{c}{0.8} & \multicolumn{2}{c}{1.0} & \multicolumn{2}{c}{1.2} \\ \hline
B_0      & -2.0   & -0.1   & -2.0   & -0.1   & -2.0   & -0.1 \\ \hline
T_s      & -0.294 & -0.838 & -0.287 & -0.882 & -0.516 & -0.889 \\
T_\theta & -0.388 & -0.868 & -0.437 & -0.924 & -0.662 & -0.932 \\
T_f      & -0.427 & -0.852 & -0.552 & -0.895 & -0.727 & -0.911 \\
\sigma   &  0.083 &  0.698 &  0.201 &  0.719 &  0.155 &  0.733 \\ 
Z_B      &  0.254 &  0.841 &  0.255 &  0.859 &  0.255 &  0.851 \\
Z_E      &  0.619 &  0.870 &  0.638 &  0.873 &  0.673 &  0.868 \\
\hline\hline
\end{array}
\end{displaymath}
\end{center}
\caption{\label{constt} 
Pearson linear correlation coefficients 
between measures of stability and the logarithm of the mean first
passage time for 15-mers simulated with a single temperature.}
\end{table}

\clearpage

\clearpage

\end{document}